\documentclass[sigplan,screen]{acmart}
\setcopyright{acmlicensed}
\usepackage{tabularx}
\acmYear{2024}
\copyrightyear{2024}
\acmConference[SenSys '24]{ACM Conference on Embedded Networked Sensor Systems}{November 4--7, 2024}{Hangzhou, China}
\acmBooktitle{ACM Conference on Embedded Networked Sensor Systems (SenSys '24), November 4--7, 2024, Hangzhou, China}
\acmDOI{10.1145/3666025.3699404}
\acmISBN{979-8-4007-0697-4/24/11}

\begin{document}

\title{Poster Abstract: LLM-Slice: Dedicated Wireless Network Slicing for Large Language Models}

\author{Boyi Liu}
\affiliation{%
  \institution{HKUST}
  \country{Hong Kong SAR}}
\email{bliubd@connect.ust.hk}
\author{Jingwen Tong}
\affiliation{%
  \institution{HKUST}
  \country{Hong Kong SAR}}
\email{eejwentong@ust.hk}
\author{Jun Zhang}
\affiliation{%
  \institution{HKUST}
  \country{Hong Kong SAR}}
\email{eejzhang@ust.hk}
\begin{abstract}
The rapid adoption of large language models (LLMs) presents new challenges for existing network architectures due to significant peak traffic and high communication uncertainty. Traditional wireless networks struggle to support efficiently, leading to intolerable response delays, disconnections, and resource wastage. To address these issues, we propose \emph{LLM-Slice}, the first system to provide dedicated communication slices for LLMs within a wireless network environment. By creating LLM-specific network slices, LLM-Slice efficiently binds services with communication resources. Based on user equipment (UE) requests and a permissions database, the system registers specific slices to offer controllable LLM services, integrating a downlink resource control module to optimize response speed, enhance resource utilization, and reduce disconnections. By deploying and validating in a real UE-gNB-CN environment, numerical results demonstrate that LLM-Slice significantly improves response speed and resource efficiency, providing a novel solution for fast and controllable LLM access in wireless networks.
\end{abstract}
\maketitle
\section{Introduction}
\begin{figure}
  \includegraphics[width=0.47\textwidth]{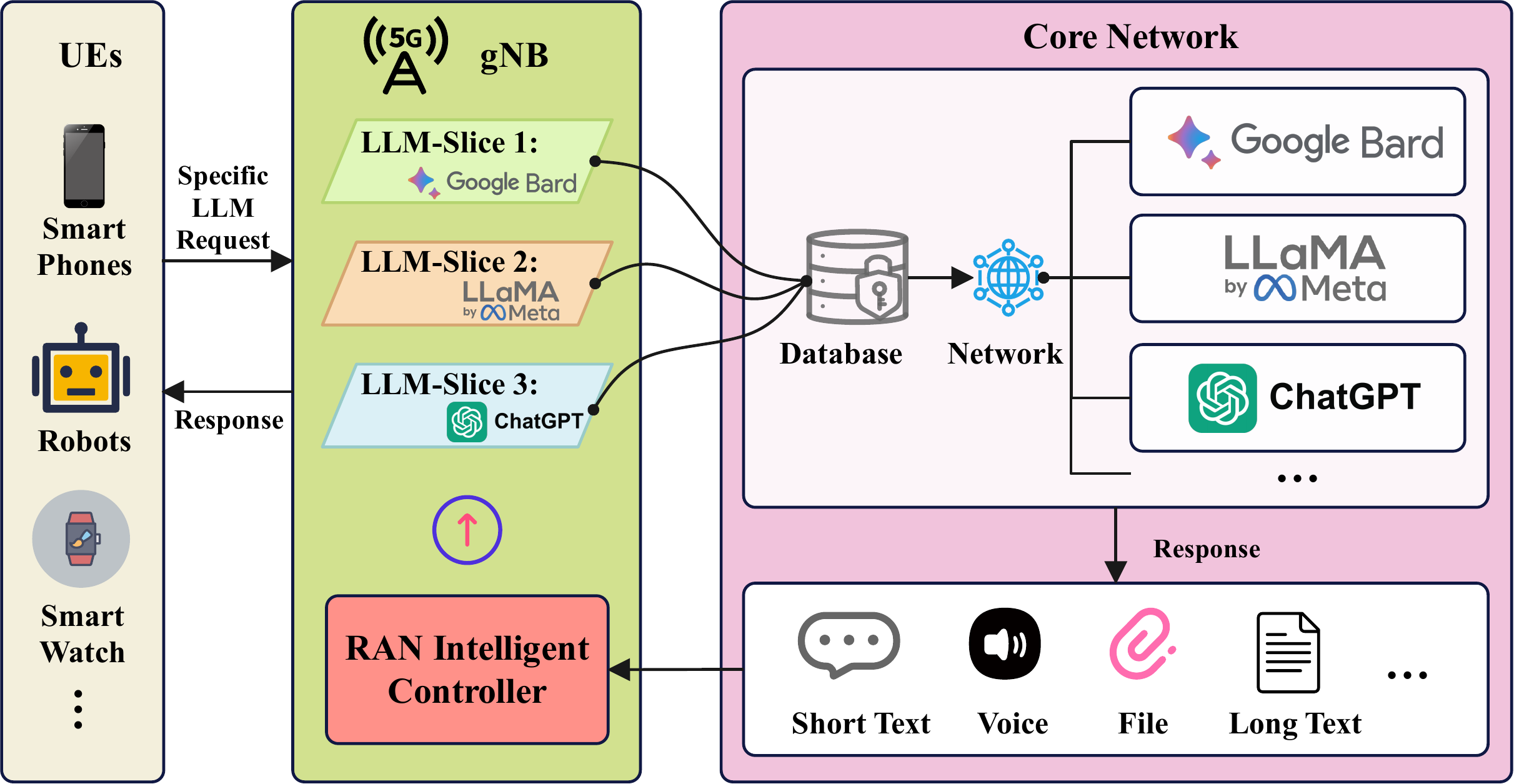}
  \caption{The Overview of the LLM-Slice System}
  \label{fig:Intro}
\end{figure}
\begin{figure*}
  \includegraphics[width=0.98\textwidth]{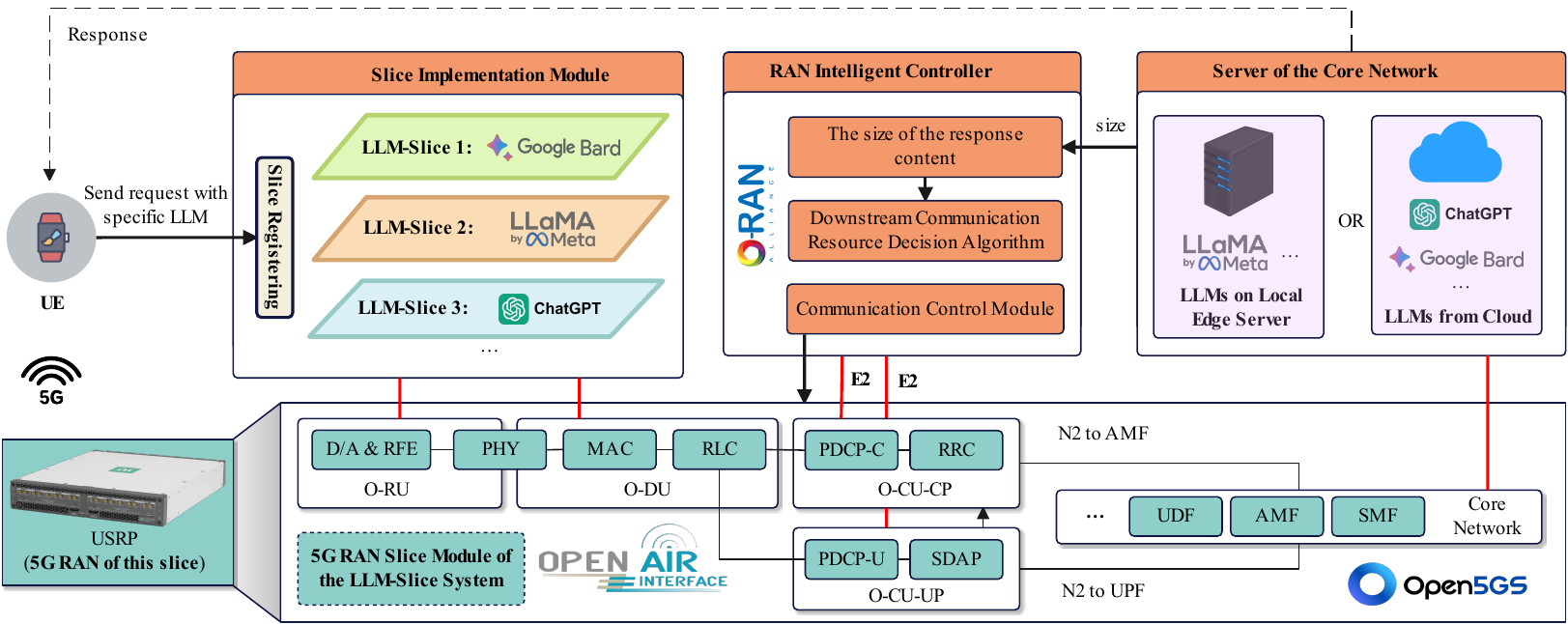}
  \caption{The Framework of the LLM-Slice System}
  \label{fig:System}
\end{figure*}
\begin{figure*}
  \includegraphics[width=0.98\textwidth]{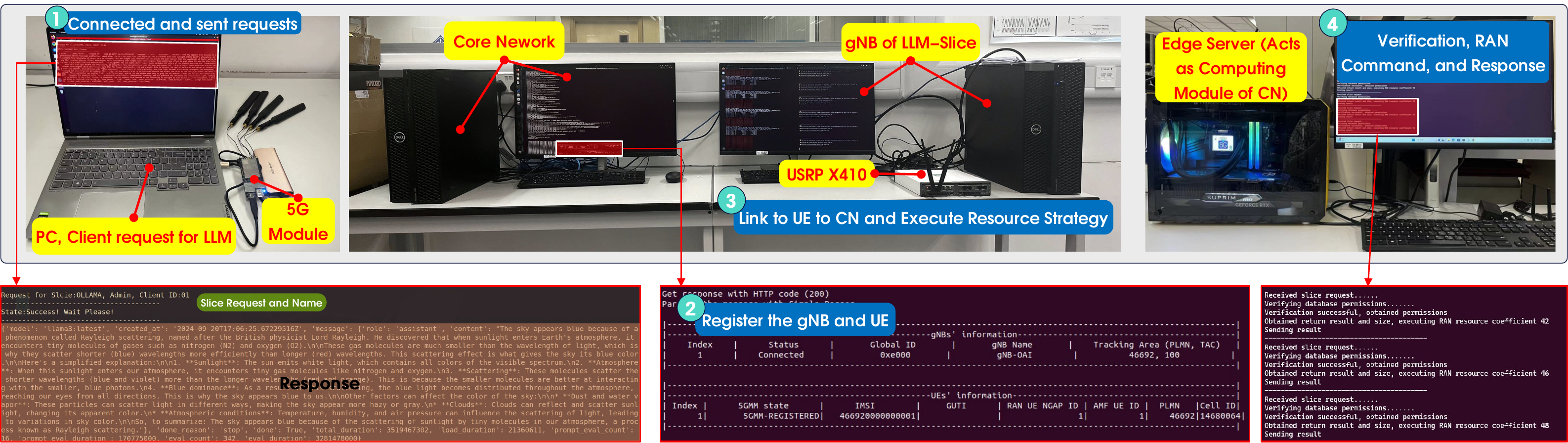}
  \caption{Real-World Implementation of LLM-Slice}
  \label{fig:Implementation}
\end{figure*}
In recent years, the rapid development and widespread application of LLMs have been revolutionizing the field of artificial intelligence \cite{chang2024survey, wang2024survey, thirunavukarasu2023large, wu2023bloomberggpt}. From natural language processing to complex problem-solving \cite{chen2023language, jackson2024natural, zhou2024large}, LLMs have demonstrated unprecedented capabilities in text understanding and generation. However, wireless communications have become one of the bottlenecks to accessing LLM services, primarily due to the special traffic patterns and resource demands of LLMs. Traditional wireless network architectures face severe challenges: 
1) LLM applications often generate \textbf{significant traffic peaks}, leading to network congestion and degradation of service quality; 
2) LLM-generated responses vary greatly in length and complexity, leading to \textbf{high downlink communication uncertainty}, making efficient allocation of network resources difficult;
3) Due to improper allocation of network resources, users may experience noticeable \textbf{response delays}, affecting user experience;
4) When processing large amounts of LLM-generated data, \textbf{downlink disconnections} may occur, resulting in information loss and service interruptions. Therefore, a lack of optimization for LLM traffic characteristics will lead to inefficient resource utilization and performance degradation.


To address these challenges, we propose and implement \emph{LLM-Slice}, the first 5G wireless network system to provide dedicated communication slices for LLMs. As shown in Figure \ref{fig:Intro}, LLM-Slice creates LLM-specific network slices to efficiently bind services and communication resources. At first, UEs request specific LLM services from gNB where LLM-Slice implements multiple dedicated slices (e.g., Google Bard, LLaMA, ChatGPT), allowing different LLM services to coexist with independent resource allocation and management. Then, the core network includes a permission database and a control module for managing user access and resource allocation. At last, a RAN Intelligent Controller (RIC) dynamically adjusts wireless resources based on LLM response characteristics, enhancing system adaptability and optimizing network resource utilization.

\section{System Design}
Figure \ref{fig:System} shows the system architecture, which consists of three fundamental components: slice implementation, RIC, and a core network server. This design enables efficient network slicing management to support various LLM services.

\textbf{Slice implementation.} This module has a multi-tier architecture with three layers: the service layer, the network function layer, and the infrastructure layer. Each layer has specific tasks that contribute to slice deployment and implementation. The service layer manages LLM-specific slices, while the network function layer handles resource allocation and the infrastructure layer provides the physical resources.

\textbf{RAN intelligent controller.} This module dynamically optimizes resource allocation based on the characteristics of LLM responses. This intelligent system analyzes content size and employs sophisticated algorithms to make real-time decisions on resource allocation, ensuring optimal network performance for LLM services. It adapts to varying LLM traffic patterns, enhancing system flexibility and efficiency.

\textbf{Core network server.} It includes a permissions database to authenticate and authorize users for specific LLM services. Additionally, it hosts the communication control module, which interfaces with the RIC to implement resource allocation decisions and manage the end-to-end network slices for LLM services.

The system operates through a coordinated workflow. Initially, the UE sends an LLM service request to the gNB. The slice implementation module then registers the appropriate LLM-specific slice based on the request. The core network server verifies user permissions and activates the slice. As the LLM generates responses, the RIC dynamically adjusts resource allocation based on response characteristics. This process ensures efficient data transmission from the LLM to the UE through the optimized network slice.
\section{System Implementation}
We implemented the LLM-Slice system on a real-world 5G testbed, as depicted in Figure \ref{fig:Implementation}. The platform consists of several key components: a host machine running the Core Network functionality, connected to an Edge Server acting as a computing module; a USRP X410 software-defined radio device serving as the gNB for LLM-Slice; and a PC equipped with a 5G module functioning as the client to send LLM requests. These components are interconnected via a 5G network, collectively realizing the functionality of LLM-Slice. 

\textbf{1) For the RAN}, we utilized a USRP X410 software-defined radio programmed with a modified version of the OpenAirInterface \cite{nikaein2014openairinterface} gNB software. We implemented LLM-specific slice support in the MAC layer and added interfaces to communicate slice information with the core network. \textbf{2) The UE}, represented by a PC with a 5G module, was configured to request LLM-specific slices. We developed a client application that sends LLM requests and measures response times. \textbf{3) For RIC}, we implemented the RIC as a separate process on the Edge Server. This controller interfaces with both the gNB and core network using the O-RAN E2 interface, which we extended to support LLM-specific metrics. \textbf{4) For LLM integration}, we deployed the LLaMA model on the Edge Server. We utilized a custom API to handle requests from the core network and return responses, which are then transmitted through the optimized network slice. The overall operating status is illustrated in Figure \ref{fig:Implementation}.
\section{Preliminary Results}
\begin{table}[h]
\centering
\caption{Comparison of LLM-Slice and Baseline 5G}
\label{tab:performance}
\begin{tabularx}{\columnwidth}{|X|c|c|c|}
\hline
\textbf{Metric} & \textbf{Baseline} & \textbf{LLM-Slice} & \textbf{Improv.} \\
\hline
Avg. Latency & 250 ms & 120 ms & 52\% \\
Resource Utilization & 65\% & 85\% & 30.8\% \\
Downlink Stability & 92\% & 99\% & 7.6\% \\
\hline
\end{tabularx}
\end{table}

The numerical results in Tabel \ref{tab:performance} show promising improvements across all metrics. LLM-Slice achieved a 52\% reduction in average response latency. Resource utilization improved by 30.8\%, indicating more efficient network management. Downlink stability also saw a notable increase, potentially reducing disconnections during LLM interactions. While these preliminary findings are encouraging, further studies are needed to evaluate LLM-Slice's performance with more samples and with different LLM workloads.

\section{Future Work}
Building on the current LLM-Slice framework, our future work will focus on several key enhancements to our previous works. First, we plan to integrate federated learning techniques to improve model performance and protect privacy, based on previous research in federated reinforcement and imitation learning \cite{liu2019lifelong, liu2019federated, liu2021peer, liu2020experiments, yan2021fedcm}. Second, to optimize resource allocation and reduce latency, we will introduce dynamic cloud-edge collaboration mechanisms inspired by elastic system architectures \cite{liu2022elasticros, liu2023roboec2}. Additionally, to safeguard sensitive LLM data, we will explore blockchain-based security measures \cite{zhang2022authros}. We will also incorporate advanced prediction methods and leverage recent advances in smart city applications and communication-adaptive systems \cite{liu2017singular, zheng2022applications, liu2024edgeloc} to develop more sophisticated resource allocation strategies for handling varying LLM workloads in complex urban environments. These improvements aim to create a more robust, efficient, and secure network slicing system for LLM services.

\section{Acknowledgement}
This work was supported by the Hong Kong Research Grants Council under the Areas of Excellence scheme grant AoE/E-601/22-R and NSFC/RGC Collaborative Research Scheme grant CRS\_HKUST603/22.
\bibliographystyle{ACM-Reference-Format}
\bibliography{sample-base}


\begin{thebibliography}{19}


\ifx \showCODEN    \undefined \def \showCODEN     #1{\unskip}     \fi
\ifx \showDOI      \undefined \def \showDOI       #1{#1}\fi
\ifx \showISBNx    \undefined \def \showISBNx     #1{\unskip}     \fi
\ifx \showISBNxiii \undefined \def \showISBNxiii  #1{\unskip}     \fi
\ifx \showISSN     \undefined \def \showISSN      #1{\unskip}     \fi
\ifx \showLCCN     \undefined \def \showLCCN      #1{\unskip}     \fi
\ifx \shownote     \undefined \def \shownote      #1{#1}          \fi
\ifx \showarticletitle \undefined \def \showarticletitle #1{#1}   \fi
\ifx \showURL      \undefined \def \showURL       {\relax}        \fi
\providecommand\bibfield[2]{#2}
\providecommand\bibinfo[2]{#2}
\providecommand\natexlab[1]{#1}
\providecommand\showeprint[2][]{arXiv:#2}

\bibitem[Chang et~al\mbox{.}(2024)]%
        {chang2024survey}
\bibfield{author}{\bibinfo{person}{Yupeng Chang}, \bibinfo{person}{Xu Wang}, \bibinfo{person}{Jindong Wang}, \bibinfo{person}{Yuan Wu}, \bibinfo{person}{Linyi Yang}, \bibinfo{person}{Kaijie Zhu}, \bibinfo{person}{Hao Chen}, \bibinfo{person}{Xiaoyuan Yi}, \bibinfo{person}{Cunxiang Wang}, \bibinfo{person}{Yidong Wang}, {et~al\mbox{.}}} \bibinfo{year}{2024}\natexlab{}.
\newblock \showarticletitle{A survey on evaluation of large language models}.
\newblock \bibinfo{journal}{\emph{ACM Transactions on Intelligent Systems and Technology}} \bibinfo{volume}{15}, \bibinfo{number}{3} (\bibinfo{year}{2024}), \bibinfo{pages}{1--45}.
\newblock


\bibitem[Chen et~al\mbox{.}(2023)]%
        {chen2023language}
\bibfield{author}{\bibinfo{person}{Xi Chen}, \bibinfo{person}{Giacomo Anerdi}, \bibinfo{person}{Daniel~Stanley Tan}, {and} \bibinfo{person}{Stefano Bromuri}.} \bibinfo{year}{2023}\natexlab{}.
\newblock \showarticletitle{Language Modeling in Logistics: Customer Calling Prediction}. In \bibinfo{booktitle}{\emph{Proceedings of the European Symposium on Artificial Neural Networks, Computational Intelligence and Machine Learning, Bruges, Belgium}}. \bibinfo{pages}{4--6}.
\newblock


\bibitem[Jackson et~al\mbox{.}(2024)]%
        {jackson2024natural}
\bibfield{author}{\bibinfo{person}{Ilya Jackson}, \bibinfo{person}{Maria Jesus~Saenz}, {and} \bibinfo{person}{Dmitry Ivanov}.} \bibinfo{year}{2024}\natexlab{}.
\newblock \showarticletitle{From natural language to simulations: applying AI to automate simulation modelling of logistics systems}.
\newblock \bibinfo{journal}{\emph{International Journal of Production Research}} \bibinfo{volume}{62}, \bibinfo{number}{4} (\bibinfo{year}{2024}), \bibinfo{pages}{1434--1457}.
\newblock


\bibitem[Liu et~al\mbox{.}(2017)]%
        {liu2017singular}
\bibfield{author}{\bibinfo{person}{Boyi Liu}, \bibinfo{person}{Jieren Cheng}, \bibinfo{person}{Kuanqi Cai}, \bibinfo{person}{Pengchao Shi}, {and} \bibinfo{person}{Xiangyan Tang}.} \bibinfo{year}{2017}\natexlab{}.
\newblock \showarticletitle{Singular point probability improve LSTM network performance for long-term traffic flow prediction}. In \bibinfo{booktitle}{\emph{Theoretical Computer Science: 35th National Conference, NCTCS 2017, Wuhan, China, October 14-15, 2017, Proceedings}}. Springer, \bibinfo{pages}{328--340}.
\newblock


\bibitem[Liu et~al\mbox{.}(2024)]%
        {liu2024edgeloc}
\bibfield{author}{\bibinfo{person}{Boyi Liu}, \bibinfo{person}{Jingwen Tong}, {and} \bibinfo{person}{Yufan Zhuang}.} \bibinfo{year}{2024}\natexlab{}.
\newblock \showarticletitle{EdgeLoc: A Communication-Adaptive Parallel System for Real-Time Localization in Infrastructure-Assisted Autonomous Driving}.
\newblock \bibinfo{journal}{\emph{arXiv preprint arXiv:2405.12120}} (\bibinfo{year}{2024}).
\newblock


\bibitem[Liu et~al\mbox{.}(2021)]%
        {liu2021peer}
\bibfield{author}{\bibinfo{person}{Boyi Liu}, \bibinfo{person}{Lujia Wang}, \bibinfo{person}{Xinquan Chen}, \bibinfo{person}{Lexiong Huang}, \bibinfo{person}{Dong Han}, {and} \bibinfo{person}{Cheng-Zhong Xu}.} \bibinfo{year}{2021}\natexlab{}.
\newblock \showarticletitle{Peer-assisted robotic learning: a data-driven collaborative learning approach for cloud robotic systems}. In \bibinfo{booktitle}{\emph{2021 IEEE international conference on robotics and automation (ICRA)}}. IEEE, \bibinfo{pages}{4062--4070}.
\newblock


\bibitem[Liu et~al\mbox{.}(2022)]%
        {liu2022elasticros}
\bibfield{author}{\bibinfo{person}{Boyi Liu}, \bibinfo{person}{Lujia Wang}, {and} \bibinfo{person}{Ming Liu}.} \bibinfo{year}{2022}\natexlab{}.
\newblock \showarticletitle{ElasticROS: An Elastically Collaborative Robot Operation System for Fog and Cloud Robotics}.
\newblock \bibinfo{journal}{\emph{arXiv preprint arXiv:2209.01774}} (\bibinfo{year}{2022}).
\newblock


\bibitem[Liu et~al\mbox{.}(2023)]%
        {liu2023roboec2}
\bibfield{author}{\bibinfo{person}{Boyi Liu}, \bibinfo{person}{Lujia Wang}, {and} \bibinfo{person}{Ming Liu}.} \bibinfo{year}{2023}\natexlab{}.
\newblock \showarticletitle{Roboec2: A novel cloud robotic system with dynamic network offloading assisted by amazon ec2}.
\newblock \bibinfo{journal}{\emph{IEEE Transactions on Automation Science and Engineering}} (\bibinfo{year}{2023}).
\newblock


\bibitem[Liu et~al\mbox{.}(2019a)]%
        {liu2019federated}
\bibfield{author}{\bibinfo{person}{Boyi Liu}, \bibinfo{person}{Lujia Wang}, \bibinfo{person}{Ming Liu}, {and} \bibinfo{person}{Cheng-Zhong Xu}.} \bibinfo{year}{2019}\natexlab{a}.
\newblock \showarticletitle{Federated Imitation Learning: A Novel Framework for Cloud Robotic Systems with Heterogeneous Sensor Data}.
\newblock \bibinfo{journal}{\emph{IEEE Robotics and Automation Letters}} \bibinfo{volume}{5}, \bibinfo{number}{2} (\bibinfo{year}{2019}), \bibinfo{pages}{3509--3516}.
\newblock


\bibitem[Liu et~al\mbox{.}(2019b)]%
        {liu2019lifelong}
\bibfield{author}{\bibinfo{person}{Boyi Liu}, \bibinfo{person}{Lujia Wang}, \bibinfo{person}{Ming Liu}, {and} \bibinfo{person}{Cheng-Zhong Xu}.} \bibinfo{year}{2019}\natexlab{b}.
\newblock \showarticletitle{Lifelong federated reinforcement learning: a learning architecture for navigation in cloud robotic systems}.
\newblock \bibinfo{journal}{\emph{IEEE Robotics and Automation Letters}} \bibinfo{volume}{4}, \bibinfo{number}{4} (\bibinfo{year}{2019}), \bibinfo{pages}{4555--4562}.
\newblock


\bibitem[Liu et~al\mbox{.}(2020)]%
        {liu2020experiments}
\bibfield{author}{\bibinfo{person}{Boyi Liu}, \bibinfo{person}{Bingjie Yan}, \bibinfo{person}{Yize Zhou}, \bibinfo{person}{Yifan Yang}, {and} \bibinfo{person}{Yixian Zhang}.} \bibinfo{year}{2020}\natexlab{}.
\newblock \showarticletitle{Experiments of federated learning for covid-19 chest x-ray images}.
\newblock \bibinfo{journal}{\emph{arXiv preprint arXiv:2007.05592}} (\bibinfo{year}{2020}).
\newblock


\bibitem[Nikaein et~al\mbox{.}(2014)]%
        {nikaein2014openairinterface}
\bibfield{author}{\bibinfo{person}{Navid Nikaein}, \bibinfo{person}{Mahesh~K Marina}, \bibinfo{person}{Saravana Manickam}, \bibinfo{person}{Alex Dawson}, \bibinfo{person}{Raymond Knopp}, {and} \bibinfo{person}{Christian Bonnet}.} \bibinfo{year}{2014}\natexlab{}.
\newblock \showarticletitle{OpenAirInterface: A flexible platform for 5G research}.
\newblock \bibinfo{journal}{\emph{ACM SIGCOMM Computer Communication Review}} \bibinfo{volume}{44}, \bibinfo{number}{5} (\bibinfo{year}{2014}), \bibinfo{pages}{33--38}.
\newblock


\bibitem[Thirunavukarasu et~al\mbox{.}(2023)]%
        {thirunavukarasu2023large}
\bibfield{author}{\bibinfo{person}{Arun~James Thirunavukarasu}, \bibinfo{person}{Darren Shu~Jeng Ting}, \bibinfo{person}{Kabilan Elangovan}, \bibinfo{person}{Laura Gutierrez}, \bibinfo{person}{Ting~Fang Tan}, {and} \bibinfo{person}{Daniel Shu~Wei Ting}.} \bibinfo{year}{2023}\natexlab{}.
\newblock \showarticletitle{Large language models in medicine}.
\newblock \bibinfo{journal}{\emph{Nature medicine}} \bibinfo{volume}{29}, \bibinfo{number}{8} (\bibinfo{year}{2023}), \bibinfo{pages}{1930--1940}.
\newblock


\bibitem[Wang et~al\mbox{.}(2024)]%
        {wang2024survey}
\bibfield{author}{\bibinfo{person}{Lei Wang}, \bibinfo{person}{Chen Ma}, \bibinfo{person}{Xueyang Feng}, \bibinfo{person}{Zeyu Zhang}, \bibinfo{person}{Hao Yang}, \bibinfo{person}{Jingsen Zhang}, \bibinfo{person}{Zhiyuan Chen}, \bibinfo{person}{Jiakai Tang}, \bibinfo{person}{Xu Chen}, \bibinfo{person}{Yankai Lin}, {et~al\mbox{.}}} \bibinfo{year}{2024}\natexlab{}.
\newblock \showarticletitle{A survey on large language model based autonomous agents}.
\newblock \bibinfo{journal}{\emph{Frontiers of Computer Science}} \bibinfo{volume}{18}, \bibinfo{number}{6} (\bibinfo{year}{2024}), \bibinfo{pages}{186345}.
\newblock


\bibitem[Wu et~al\mbox{.}(2023)]%
        {wu2023bloomberggpt}
\bibfield{author}{\bibinfo{person}{Shijie Wu}, \bibinfo{person}{Ozan Irsoy}, \bibinfo{person}{Steven Lu}, \bibinfo{person}{Vadim Dabravolski}, \bibinfo{person}{Mark Dredze}, \bibinfo{person}{Sebastian Gehrmann}, \bibinfo{person}{Prabhanjan Kambadur}, \bibinfo{person}{David Rosenberg}, {and} \bibinfo{person}{Gideon Mann}.} \bibinfo{year}{2023}\natexlab{}.
\newblock \showarticletitle{Bloomberggpt: A large language model for finance}.
\newblock \bibinfo{journal}{\emph{arXiv preprint arXiv:2303.17564}} (\bibinfo{year}{2023}).
\newblock


\bibitem[Yan et~al\mbox{.}(2021)]%
        {yan2021fedcm}
\bibfield{author}{\bibinfo{person}{Bingjie Yan}, \bibinfo{person}{Boyi Liu}, \bibinfo{person}{Lujia Wang}, \bibinfo{person}{Yize Zhou}, \bibinfo{person}{Zhixuan Liang}, \bibinfo{person}{Ming Liu}, {and} \bibinfo{person}{Cheng-Zhong Xu}.} \bibinfo{year}{2021}\natexlab{}.
\newblock \showarticletitle{Fedcm: A real-time contribution measurement method for participants in federated learning}. In \bibinfo{booktitle}{\emph{2021 International joint conference on neural networks (IJCNN)}}. IEEE, \bibinfo{pages}{1--8}.
\newblock


\bibitem[Zhang et~al\mbox{.}(2022)]%
        {zhang2022authros}
\bibfield{author}{\bibinfo{person}{Shenhui Zhang}, \bibinfo{person}{Wenkai Li}, \bibinfo{person}{Xiaoqi Li}, {and} \bibinfo{person}{Boyi Liu}.} \bibinfo{year}{2022}\natexlab{}.
\newblock \showarticletitle{Authros: Secure data sharing among robot operating systems based on ethereum}. In \bibinfo{booktitle}{\emph{2022 IEEE 22nd International Conference on Software Quality, Reliability and Security (QRS)}}. IEEE, \bibinfo{pages}{147--156}.
\newblock


\bibitem[Zheng et~al\mbox{.}(2022)]%
        {zheng2022applications}
\bibfield{author}{\bibinfo{person}{Zhaohua Zheng}, \bibinfo{person}{Yize Zhou}, \bibinfo{person}{Yilong Sun}, \bibinfo{person}{Zhang Wang}, \bibinfo{person}{Boyi Liu}, {and} \bibinfo{person}{Keqiu Li}.} \bibinfo{year}{2022}\natexlab{}.
\newblock \showarticletitle{Applications of federated learning in smart cities: recent advances, taxonomy, and open challenges}.
\newblock \bibinfo{journal}{\emph{Connection Science}} \bibinfo{volume}{34}, \bibinfo{number}{1} (\bibinfo{year}{2022}), \bibinfo{pages}{1--28}.
\newblock


\bibitem[Zhou et~al\mbox{.}(2024)]%
        {zhou2024large}
\bibfield{author}{\bibinfo{person}{Hao Zhou}, \bibinfo{person}{Chengming Hu}, \bibinfo{person}{Ye Yuan}, \bibinfo{person}{Yufei Cui}, \bibinfo{person}{Yili Jin}, \bibinfo{person}{Can Chen}, \bibinfo{person}{Haolun Wu}, \bibinfo{person}{Dun Yuan}, \bibinfo{person}{Li Jiang}, \bibinfo{person}{Di Wu}, {et~al\mbox{.}}} \bibinfo{year}{2024}\natexlab{}.
\newblock \showarticletitle{Large language model (llm) for telecommunications: A comprehensive survey on principles, key techniques, and opportunities}.
\newblock \bibinfo{journal}{\emph{arXiv preprint arXiv:2405.10825}} (\bibinfo{year}{2024}).
\newblock


\end{thebibliography}

\end{document}